\documentclass[preprint,prX]{revtex4}%
\usepackage{amsfonts}
\usepackage{amsmath}
\usepackage{amssymb}
\usepackage{graphicx}
\usepackage{bm}%
\begin{document}
\title[Fermi liquid effects on the effective vector vertex]{Superfluid response and the neutrino emissivity of baryon matter: Fermi-liquid effects.}
\author{L. B. Leinson}
\affiliation{Institute of Terrestrial Magnetism, Ionosphere and Radio Wave Propagation RAS,
RU-142190 Troitsk, Moscow Region, Russia}

\begin{abstract}
The linear response of a nonrelativistic superfluid baryon system on an
external weak field is investigated while taking into account of the
Fermi-liquid interactions. We generalize the theory developed by Leggett for a
superfluid Fermi-liquid at finite temperature to the case of timelike momentum
transfer typical of the problem of neutrino emission from neutron stars. A
space-like kinematics is also analysed for completeness and compared with
known results.

We use the obtained response functions to derive the neutrino energy losses
caused by recombination of broken pairs in the electrically neutral superfluid
baryon matter. We find that the dominant neutrino radiation occurs through the
axial-vector neutral currents. The emissivity is found to be of the same order
as in the BCS approximation, but the details of its temperature dependence are
modified by the Fermi-liquid interactions.

The role of electromagnetic correlations in the pairing case of protons
interacting with the electron background is discussed in the conclusion.

\end{abstract}
\startpage{1}
\maketitle

\section{Introduction}

Thermal excitations in superfluid baryon matter of neutron stars, in the form
of broken Cooper pairs, can recombine into the condensate by emitting neutrino
pairs via neutral weak currents. This process was suggested \cite{FRS76} many
years ago as an efficient mechanism for cooling of neutron stars in some
ranges of temperature and/or matter density. The interest in this process has
been recently revived \cite{LP06}--\cite{SR} in connection with the fact that
the existing theory of thermal neutrino radiation from superfluid neutron
matter leads to a rapid cooling of the neutron star crust, which is in
dramatic discrepancy with the observed data of superbursts \cite{Cumming},
\cite{Gupta}. It was realized that a better understanding of an efficiency of
the neutrino emission in the pair recombination is necessary to explain modern observations.

The relevant input for calculation of neutrino energy losses from the medium
is the imaginary part of the retarded weak-polarization tensor intimately
connected with the autocorrelation function of weak currents in the medium.
Though the theoretical investigation of the autocorrelation functions of
strong-interacting superfluid fermions was started more than four decades ago
the complete theory of the problem does not yet exist. Leggett's theory of a
superfluid Fermi liquid \cite{Leggett} is limited to the case when both the
transferred energy and momentum are small compared to the superfluid energy
gap, i.e., $\omega,\mathrm{q}\ll\Delta$. This theory cannot be applied to
calculations of neutrino energy losses because, in this case, we need the
medium response onto an external neutrino field in the time-like kinematic
domain, $\omega>\mathrm{q}$, and $\omega>2\Delta$, as required by the total
energy $\omega=\omega_{1}+\omega_{2}$ and momentum $\mathbf{q=q}%
_{1}+\mathbf{q}_{2}$ of escaping neutrino pair.

The well-known Larkin-Migdal theory \cite{Larkin} is restricted to the case of
zero temperature. Recently, the calculation of the neutrino energy losses was
undertaken in Refs. \cite{KV08}, \cite{SR}, where the imaginary part of the
autocorrelation functions was calculated for a superfluid neutron matter at
zero temperature. This approach is apparently inconsistent because the
imaginary part of retarded polarization functions substantially depends on the
temperature [see Eqs. (\ref{L}), (\ref{T}), (\ref{ImK00}), and (\ref{ImKij})
of this work]. One more inconsistency of the work \cite{SR} is, that the
temporal component of the axial-vector current cannot be discarded, as it is
done by the authors. This relativistic correction contributes to the neutrino
energy losses of the same order as the spin-density fluctuations, i.e.
$\propto V_{F}^{2}$. This was pointed out for the first time in Ref.
\cite{Kaminker}. Below, we will return to the discussion of these works and
compare our result with that obtained in Refs. \cite{KV08}, \cite{SR} and in
some earlier works.

The appropriate, temperature-dependent approach is developed in Ref.
\cite{L08}, where the mean-field BCS approximation is used to calculate the
superfluid response in the vector channel. To include the Fermi-liquid effects
discarded in the BCS approximation, in this paper, we first generalize
Leggett's theory to the case of arbitrary momentum transfer. We evaluate the
weak-interaction effective vertices and the autocorrelation functions while
taking into account strong residual particle-hole interactions. To obtain a
solution of Leggett's equations in reasonably simple form, we approximate the
particle-hole interactions by its first two harmonics with the aid of the
usual Landau parameters. Within these constraints we obtain the general
expression for the autocorrelation functions and then focus on the superfluid
response in the time-like kinematic domain. We investigate both the vector
channel and the axial channel of weak interactions to evaluate the rate of
neutrino energy loss through neutral weak currents caused by recombination of
electrically neutral baryons.

The role of electromagnetic correlations in the pairing case of charged
baryons interacting with the electron background deserves a separate
consideration. The quantum transitions of charged quasiparticles can excite
background electrons, thus inducing the neutrino-pair emission by the electron
plasma \cite{L00}, \cite{L01}. In summary, we briefly discuss this problem in
the light of modern theory to understand whether the plasma effects can lead
to noticeable neutrino energy losses through the vector channel.

The paper is organized as follows. The next section contains some preliminary
notes and outlines some important properties of Green's functions and the
one-loop integrals used below. In Sec. III, we discuss the set of equations
derived by Leggett for calculation of correlation functions of a superfluid
Fermi liquid at finite temperature. In Sec. IV, we consider the superfluid
response in the vector channel. Because of the conservation of the vector weak
current it is sufficient to consider only the longitudinal and transverse
autocorrelation functions. The correlation functions in the axial channel are
evaluated in Sec. V. As an application of our findings, in Sec. VI, we
evaluate neutrino energy losses through neutral weak currents caused by the
pair recombination in superfluid neutron matter. Some numerical estimates of
the neutrino energy losses are represented in Sec. VII. Section VIII contains
a short summary of our findings and the conclusion.

In this work we use the standard model of weak interactions, the system of
units $\hbar=c=1,$ and the Boltzmann constant $k_{B}=1$.

\section{Preliminary notes and notation}

In our analysis, we will use the fact that the Fermi-liquid interactions do
not interfere with the pairing phenomenon if approximate hole-particle
symmetry is maintained in the system; i.e. the Fermi-liquid interactions
remain unchanged upon pairing. According to Landau's theory, near the Fermi
surface, $\mathsf{p\simeq p}^{\prime}\mathsf{\simeq p}_{F}$, the Fermi-liquid
interactions can be reduced to the interactions in the particle-hole channel.
We will assume that the effective interaction amplitude is the function of the
angle between incoming momenta $\mathbf{p}$ and $\mathbf{p}^{\prime}$ and can
be parametrized as the sum of the scalar and exchange terms%
\begin{equation}
a^{2}\rho\hat{\Gamma}^{\omega}\left(  \mathbf{nn}^{\prime}\right)  =f\left(
\mathbf{nn}^{\prime}\right)  +g\left(  \mathbf{nn}^{\prime}\right)
{\textstyle\sum_{i}}
\hat{\sigma}_{i}\hat{\sigma}_{i}^{\prime}. \label{phin}%
\end{equation}
Here and below, $\rho=\mathrm{p}_{F}M^{\ast}/\pi^{2}$ is the density of states
near the Fermi surface; $\mathbf{n}=\mathbf{p/}\mathsf{p}$ and $\mathbf{n}%
^{\prime}=\mathbf{p}^{\prime}\mathbf{/}\mathsf{p}^{\prime}$ are the unit
vectors specifying directions of incoming momenta, $a\simeq1$ is a usual
Green's function renormalization constant independent of $\omega,\mathbf{q}$,
and $T$, and $\hat{\sigma}_{i}$ ($i=1,2,3$) stand for Pauli spin matrices. The
pairing interaction, irreducible in the channel of two quasiparticles, is
renormalized in the same manner%
\begin{equation}
a^{2}\rho\hat{\Gamma}^{\varphi}\left(  \mathbf{nn}^{\prime}\right)
=\Gamma_{a}^{\varphi}\left(  \mathbf{nn}^{\prime}\right)  +\Gamma_{b}%
^{\varphi}\left(  \mathbf{nn}^{\prime}\right)
{\textstyle\sum_{i}}
\hat{\sigma}_{i}\hat{\sigma}_{i}^{\prime}. \label{fksi}%
\end{equation}
We will consider the case when the pairing occurs only between two
quasiparticles with the total spin $S=0$. Then the irreducible pairing
amplitude is to be taken as the singlet,%
\begin{equation}
a^{2}\rho\hat{\Gamma}^{\varphi}\left(  \mathbf{nn}^{\prime}\right)
\rightarrow\Gamma^{\varphi}\left(  \mathbf{nn}^{\prime}\right)  \equiv
\Gamma_{a}^{\varphi}\left(  \mathbf{nn}^{\prime}\right)  -3\Gamma_{b}%
^{\varphi}\left(  \mathbf{nn}^{\prime}\right)  . \label{Samp}%
\end{equation}

Since the baryonic component of stellar matter is in thermal equilibrium at
some temperature $T$, we adopt the Matsubara Green's functions for the
description of the superfluid condensate and for evaluation of the
polarization tensor. In the case of $^{1}S_{0}$ pairing, near the Fermi
surface, these are given by: \cite{AGD}:%
\begin{equation}
G\left(  p_{n},\mathbf{p}\right)  =a\frac{-ip_{n}-\varepsilon_{\mathbf{p}}%
}{p_{n}^{2}+E_{\mathbf{p}}^{2}},\ G_{h}\left(  p_{n},\mathbf{p}\right)
=a\frac{ip_{n}-\varepsilon_{\mathbf{p}}}{p_{n}^{2}+E_{\mathbf{p}}^{2}%
},\ F\left(  p_{n},\mathbf{p}\right)  =a\frac{\Delta}{p_{n}^{2}+E_{\mathbf{p}%
}^{2}}, \label{GF}%
\end{equation}
where $p_{n}=\pi\left(  2n+1\right)  T$ with $n=0,\pm1,\pm2,...$ is the
fermionic Matsubara frequency. In the above equation, $G$ and $G_{h}$
represent the propagators of a particle and of a hole, respectively, and $F$
is the anomalous propagator, i.e. the amplitude of the quasiparticle
transition into a hole and a correlated pair. For the inverse process:
$F^{\dagger}\left(  p_{n},\mathbf{p}\right)  =F\left(  p_{n},\mathbf{p}%
\right)  $.

We use the momentum representation and the following notation
\begin{equation}
\varepsilon_{\mathbf{p}}=\frac{\mathsf{p}^{2}}{2M^{\ast}}-\frac{\mathsf{p}%
_{F}^{2}}{2M^{\ast}}\simeq\frac{\mathsf{p}_{F}}{M^{\ast}}(\mathsf{p}%
-\mathsf{p}_{F}), \label{ksi}%
\end{equation}
where $M^{\ast}=p_{F}/V_{F}$ is the effective mass of a quasiparticle, and the
energy of a quasiparticle is
\begin{equation}
E_{\mathbf{p}}=\sqrt{\varepsilon_{\mathbf{p}}^{2}+\Delta^{2}\left(  T\right)
}. \label{eps}%
\end{equation}

We designate as $L_{X,X}\left(  \omega,\mathbf{q;p}\right)  $ the analytical
continuation onto the upper-half plane of complex variable $\omega$ of the
following Matsubara sums:%
\begin{equation}
L_{XX^{\prime}}\left(  \omega_{m},\mathbf{p+}\frac{\mathbf{q}}{2}%
\mathbf{;p-}\frac{\mathbf{q}}{2}\right)  =T\sum_{p_{n}}X\left(  p_{n}%
+\omega_{m},\mathbf{p+}\frac{\mathbf{q}}{2}\right)  X^{\prime}\left(
p_{n},\mathbf{p-}\frac{\mathbf{q}}{2}\right)  , \label{LXX}%
\end{equation}
where $X,X^{\prime}\in G,F,G^{h}$.

In Leggett's equations, which we are going to exploit, the spin dependence is
already taken into account, and $\sum_{\mathbf{p,\sigma}}$ \ is everywhere
replaced by $2\sum_{\mathbf{p}}$. It is convenient to divide the integration
over the momentum space into the integration over the solid angle and over the
energy according%
\begin{equation}
\int\frac{2d^{3}\mathrm{p}}{\left(  2\pi\right)  ^{3}}\cdot\cdot\cdot=\rho
\int\frac{d\mathbf{n}}{4\pi}\int_{-\infty}^{\infty}d\varepsilon_{\mathbf{p}%
}\cdot\cdot\cdot\label{1}%
\end{equation}
and operate with integrals over the quasiparticle energy:%
\begin{equation}
\mathcal{I}_{XX^{\prime}}\left(  \omega,\mathrm{q}\cos\theta,T\right)
\equiv\int_{-\infty}^{\infty}d\varepsilon_{\mathbf{p}}L_{XX^{\prime}}\left(
\omega,\mathbf{p+}\frac{\mathbf{q}}{2}\mathbf{,p-}\frac{\mathbf{q}}{2}\right)
. \label{IXX}%
\end{equation}
These are functions of $\omega$, $\mathrm{q,}$ and $\cos\theta=\mathbf{nn}%
_{\mathbf{q}}$, which is the polar angle of the direction of the momentum
$\mathbf{p}=\mathrm{p}\mathbf{n}$ relative to the direction of $\mathbf{n}%
_{\mathbf{q}}=\mathbf{q}/\mathrm{q}$ as the $z$ axis.

The functions $\mathcal{I}_{XX^{\prime}}$ possess the following properties,
which can be derived by a straightforward calculation \cite{Leggett}$:$
\begin{equation}
\mathcal{I}_{GF}=-\mathcal{I}_{FG},\ \mathcal{I}_{FG_{h}}=-\mathcal{I}%
_{G_{h}F}, \label{fg}%
\end{equation}%
\begin{equation}
\mathcal{I}_{G_{h}F}+\mathcal{I}_{FG}=\frac{\omega}{\Delta}\mathcal{I}_{FF},
\label{gf}%
\end{equation}%
\begin{equation}
\mathcal{I}_{G_{h}F}-\mathcal{I}_{FG}=-\frac{\mathbf{qv}}{\Delta}%
\mathcal{I}_{FF}, \label{ff1}%
\end{equation}%
\begin{equation}
-\left(  \mathcal{I}_{GG_{h}}+\mathcal{I}_{FF}\right)  =A_{0}+\frac{\left(
\mathbf{qv}\right)  ^{2}-\omega^{2}}{2\Delta^{2}}\mathcal{I}_{FF}. \label{2}%
\end{equation}
Here $\mathbf{v=}V_{F}\mathbf{n}$, and the quantity $A_{0}=-\left(
\mathcal{I}_{GG_{h}}+\mathcal{I}_{FF}\right)  _{\mathbf{q}=0,\omega=0}$
satisfies the gap equation%
\begin{equation}
1-\Gamma_{0}^{\varphi}A_{0}=0, \label{GAPEQ}%
\end{equation}
where $\Gamma_{0}^{\varphi}$ is the zeroth harmonic of the singlet pairing
amplitude (\ref{Samp}).

The key role in the medium response theory belongs to the functions defined as
the following combinations of the above loop integrals:%
\begin{equation}
\lambda\left(  \omega,\mathbf{qn}\right)  \equiv a^{-2}\mathcal{I}_{FF},
\label{lam}%
\end{equation}%
\begin{equation}
\kappa\left(  \omega,\mathbf{qn}\right)  \equiv a^{-2}\left(  \frac{1}%
{2}\left(  \mathcal{I}_{GG}+\mathcal{I}_{G^{h}G^{h}}\right)  +\mathcal{I}%
_{FF}\right)  , \label{kap}%
\end{equation}%
\begin{equation}
\chi\left(  \omega,\mathbf{qn}\right)  \equiv a^{-2}\frac{1}{2}\left(
\mathcal{I}_{GG}-\mathcal{I}_{G^{h}G^{h}}\right)  . \label{khi}%
\end{equation}
These can be derived in the following form:%
\begin{equation}
\lambda=-\frac{\Delta^{2}}{4}\int_{-\infty}^{\infty}\frac{d\varepsilon
_{\mathbf{p}}}{E_{+}E_{-}}\left[  \left(  E_{\mathbf{+}}+E_{\mathbf{-}%
}\right)  \Phi_{+}-\left(  E_{\mathbf{+}}-E_{\mathbf{-}}\right)  \Phi
_{-}\right]  , \label{l}%
\end{equation}%
\begin{equation}
\kappa=\frac{\mathbf{qv}}{4}\int_{-\infty}^{\infty}\frac{d\varepsilon
_{\mathbf{p}}}{E_{+}E_{-}}\left[  \left(  E_{\mathbf{-}}\varepsilon
_{\mathbf{+}}-\varepsilon_{\mathbf{-}}E_{\mathbf{+}}\right)  \Phi_{+}+\left(
\varepsilon_{\mathbf{-}}E_{\mathbf{+}}+E_{\mathbf{-}}\varepsilon_{\mathbf{+}%
}\right)  \Phi_{-}\right]  , \label{k}%
\end{equation}%
\begin{equation}
\chi=\frac{\omega}{4}\int_{-\infty}^{\infty}\frac{d\varepsilon_{\mathbf{p}}%
}{E_{+}E_{-}}\left[  \left(  E_{\mathbf{-}}\varepsilon_{\mathbf{+}%
}-\varepsilon_{\mathbf{-}}E_{\mathbf{+}}\right)  \Phi_{+}+\left(
\varepsilon_{\mathbf{-}}E_{\mathbf{+}}+E_{\mathbf{-}}\varepsilon_{\mathbf{+}%
}\right)  \Phi_{-}\right]  . \label{kh}%
\end{equation}
To shorten the expressions, we use the following notation:%
\begin{equation}
\varepsilon_{\pm}\equiv\varepsilon_{\mathbf{p\pm}\frac{\mathbf{q}}{2}%
},\ E_{\pm}\equiv E_{\mathbf{p\pm}\frac{\mathbf{q}}{2}}, \label{3}%
\end{equation}
and
\begin{equation}
\Phi_{\pm}=\frac{1}{\left(  \omega+i0\right)  ^{2}-\left(  E_{\mathbf{+}}\pm
E_{\mathbf{-}}\right)  ^{2}}\left(  \tanh\frac{E_{\mathbf{+}}}{2T}\pm
\tanh\frac{E_{\mathbf{-}}}{2T}\right)  . \label{4}%
\end{equation}

It is straightforward to verify that%
\begin{equation}
\lambda\left(  \omega,\mathbf{qn}\right)  =\lambda\left(  \omega
,-\mathbf{qn}\right)  ,\ \kappa\left(  \omega,\mathbf{qn}\right)
=\kappa\left(  \omega,-\mathbf{qn}\right)  , \label{parity}%
\end{equation}
and that the functions $\kappa\left(  \omega,\mathbf{qn}\right)  $ and
$\chi\left(  \omega,\mathbf{qn}\right)  $ are not independent because
\begin{equation}
\omega\kappa=\mathbf{qv}\chi. \label{5}%
\end{equation}

\section{Leggett's finite-temperature formalism}

The two-particle autocorrelation function is defined as
\begin{equation}
K_{\xi}\left(  \omega,\mathbf{q}\right)  \equiv\sum_{\mathbf{pp}^{\prime
},\sigma\sigma^{\prime}}\xi\left(  \mathbf{p,}\sigma\right)  \left\langle
\left\langle a_{\mathbf{p+q/}2\mathbf{,}\sigma}^{\dagger}a_{\mathbf{p-q/}%
2\mathbf{,}\sigma}:a_{\mathbf{p}^{\prime}\mathbf{-q/}2\mathbf{,}\sigma
^{\prime}}^{\dagger}a_{\mathbf{p}^{\prime}\mathbf{+q/}2\mathbf{,}%
\sigma^{\prime}}\right\rangle \right\rangle _{\omega}\xi\left(  \mathbf{p}%
^{\prime}\mathbf{,}\sigma^{\prime}\right)  , \label{6}%
\end{equation}
where $\xi\left(  \mathbf{p,}\sigma\right)  $ is a three-point vertex
responsible for the interaction of a free particle with the weak external
field. It is some function of the momentum $\mathbf{p}$ and spin
variables\textbf{ }$\sigma$; $\left\langle \left\langle A:B\right\rangle
\right\rangle _{\omega}$ is the Fourier transform of a retarded two-particle
Green's function.

The analytic form of the autocorrelation function can be immediately written,
if we know the effective (full) three-point vertices defined via the linear
correction to the quasiparticle self-energy $\Xi^{\left(  1\right)  }\left(
V\right)  $ in the external field $V$ (see, e.g., Ref. \cite{Migdal}):%

\begin{equation}
\mathcal{T}=\xi\left(  \mathbf{p,}\sigma\right)  +\frac{\partial\Xi^{\left(
1\right)  }}{\partial V}. \label{gtilde}%
\end{equation}
Near the Fermi surface, these vertices can be treated as functions of
transferred energy and momentum, $q=\left(  \omega,\mathbf{q}\right)  $, and
the direction of nucleon motion $\mathbf{n}$.

In superfluids, we have to distinguish the vertices of a particle and a hole,
which are related as $\xi_{h}\left(  \mathbf{p,}\sigma\right)  =\xi\left(
-\mathbf{p,-}\sigma\right)  $. Since there are two possible cases, $\xi\left(
-\mathbf{p,-}\sigma\right)  =\pm\xi\left(  \mathbf{p,}\sigma\right)  $, it is
convenient to consider the "even" and "odd" bare vertices%
\begin{equation}
\xi_{\pm}\left(  \mathbf{n}\right)  =\frac{1}{2}\left(  \xi\left(
\mathbf{p,}\sigma\right)  \pm\xi\left(  -\mathbf{p,-}\sigma\right)  \right)  .
\label{7}%
\end{equation}
We denote as
\begin{equation}
\mathcal{T}_{\pm}\left(  \mathbf{n}\right)  =\frac{1}{2}\left(  \mathcal{T}%
\left(  \mathbf{p,}\sigma\right)  \pm\mathcal{T}\left(  -\mathbf{p,-}%
\sigma\right)  \right)  \label{8}%
\end{equation}
the corresponding full vertices taking into account the polarization of
superfluid Fermi liquid under the influence of the external field.

In Eq. (\ref{gtilde}), the quasiparticle self-energy consists of the normal
part and the anomalous part caused by the pair condensation. In the case of
$^{1}S_{0}$ pairing, the anomalous self-energy is sensitive only to the
longitudinal vector fields, because the only kind of motion possible for the
condensate is potential flow, i.e., a density fluctuation \cite{Bogoliubov}.
Therefore for the longitudinal currents, along with the ordinary vertices
$\mathcal{T}_{\pm}$, it is necessary to consider the anomalous vertex
$\mathcal{\tilde{T}}$, responsible for excitations of the condensate.

As was derived by Leggett (see Eqs. (22) and (23) of Ref. \cite{Leggett}), the
longitudinal effective vertices $\mathcal{T}_{\pm},\mathcal{\tilde{T}}$ are to
be found from the following equations (we omit for brevity the dependence of
functions on $\omega$ and $\mathbf{q}$\textbf{)}:%

\begin{gather}
\mathcal{\tilde{T}}\left(  \mathbf{n}\right)  -\int\frac{d\mathbf{n}^{\prime}%
}{4\pi}\Gamma^{\varphi}\left(  \mathbf{nn}^{\prime}\right)  A_{0}%
\mathcal{\tilde{T}}\left(  \mathbf{n}^{\prime}\right)  -\int\frac
{d\mathbf{n}^{\prime}}{4\pi}\Gamma^{\varphi}\left(  \mathbf{nn}^{\prime
}\right)  \frac{\left(  \mathbf{qv}^{\prime}\right)  ^{2}-\omega^{2}}%
{2\Delta^{2}}\lambda\left(  \mathbf{n}^{\prime}\right)  \mathcal{\tilde{T}%
}\left(  \mathbf{n}^{\prime}\right) \nonumber\\
+\int\frac{d\mathbf{n}^{\prime}}{4\pi}\Gamma^{\varphi}\left(  \mathbf{nn}%
^{\prime}\right)  \frac{\mathbf{qv}^{\prime}}{\Delta}\lambda\left(
\mathbf{n}^{\prime}\right)  \mathcal{T}_{-}\left(  \mathbf{n}^{\prime}\right)
-\frac{\omega}{\Delta}\int\frac{d\mathbf{n}^{\prime}}{4\pi}\Gamma^{\varphi
}\left(  \mathbf{nn}^{\prime}\right)  \lambda\left(  \mathbf{n}^{\prime
}\right)  \mathcal{T}_{+}\left(  \mathbf{n}^{\prime}\right)  =0, \label{psi1}%
\end{gather}%
\begin{gather}
\mathcal{T}_{-}\left(  \mathbf{n}\right)  +\int\frac{d\mathbf{n}^{\prime}%
}{4\pi}f\left(  \mathbf{nn}^{\prime}\right)  \frac{\mathbf{qv}^{\prime}%
}{\Delta}\lambda\left(  \mathbf{n}^{\prime}\right)  \mathcal{\tilde{T}}\left(
\mathbf{n}^{\prime}\right)  -\int\frac{d\mathbf{n}^{\prime}}{4\pi}f\left(
\mathbf{nn}^{\prime}\right)  \kappa\left(  \mathbf{n}^{\prime}\right)
\mathcal{T}_{-}\left(  \mathbf{n}^{\prime}\right) \nonumber\\
+\int\frac{d\mathbf{n}^{\prime}}{4\pi}f\left(  \mathbf{nn}^{\prime}\right)
\frac{\omega}{\mathbf{qv}^{\prime}}\kappa\left(  \mathbf{n}^{\prime}\right)
\mathcal{T}_{+}\left(  \mathbf{n}^{\prime}\right)  =\xi_{-}\left(
\mathbf{n}\right)  , \label{psi2}%
\end{gather}%
\begin{gather}
\mathcal{T}_{+}\left(  \mathbf{n}\right)  -\frac{\mathrm{\omega}}{\Delta}%
\int\frac{d\mathbf{n}^{\prime}}{4\pi}f\left(  \mathbf{nn}^{\prime}\right)
\lambda\left(  \mathbf{n}^{\prime}\right)  \mathcal{\tilde{T}}\left(
\mathbf{n}^{\prime}\right)  +\int\frac{d\mathbf{n}^{\prime}}{4\pi}f\left(
\mathbf{nn}^{\prime}\right)  \frac{\omega}{\mathbf{qv}^{\prime}}\kappa\left(
\mathbf{n}^{\prime}\right)  \mathcal{T}_{-}\left(  \mathbf{n}^{\prime}\right)
\nonumber\\
-\int\frac{d\mathbf{n}^{\prime}}{4\pi}f\left(  \mathbf{nn}^{\prime}\right)
\left(  \kappa\left(  \mathbf{n}^{\prime}\right)  -2\lambda\left(
\mathbf{n}^{\prime}\right)  \right)  \mathcal{T}_{+}\left(  \mathbf{n}%
^{\prime}\right)  =\xi_{+}\left(  \mathbf{n}\right)  . \label{psi3}%
\end{gather}
In Eq. (\ref{psi1}), the irreducible pairing amplitude is to be taken as the
singlet, as given by Eq. (\ref{Samp}).

Once the effective vertices are calculated, the two-particle autocorrelation
function can be immediately found using the expressions:%
\begin{equation}
K_{\xi}=\rho\int\frac{d\mathbf{n}}{4\pi}\xi_{+}\left(  \mathbf{n}\right)
\left[  \frac{\omega}{\Delta}\lambda\left(  \mathbf{n}\right)  \mathcal{\tilde
{T}}\left(  \mathbf{n}\right)  +\frac{\omega}{\mathbf{qv}}\kappa\left(
\mathbf{n}\right)  \mathcal{T}_{-}\left(  \mathbf{n}\right)  +\left(
\kappa\left(  \mathbf{n}\right)  -2\lambda\left(  \mathbf{n}\right)  \right)
\mathcal{T}_{+}\left(  \mathbf{n}\right)  \right]  , \label{Kp}%
\end{equation}
if $\xi=\xi_{+}$, and%
\begin{equation}
K_{\xi}=\rho\int\frac{d\mathbf{n}}{4\pi}\xi_{-}\left(  \mathbf{n}\right)
\left[  -\frac{\mathbf{qv}}{\Delta}\lambda\left(  \mathbf{n}\right)
\mathcal{\tilde{T}}\left(  \mathbf{n}\right)  +\kappa\left(  \mathbf{n}%
\right)  \mathcal{T}_{-}\left(  \mathbf{n}\right)  -\frac{\omega}{\mathbf{qv}%
}\kappa\left(  \mathbf{n}\right)  \mathcal{T}_{+}\left(  \mathbf{n}\right)
\right]  \label{Km}%
\end{equation}
if $\xi=\xi_{-}$.

One can easily verify that these equations represent a generalization for the
case of finite temperatures of the Larkin-Migdal \cite{Larkin} equations
derived in the ladder approximation for the vertices modified by strong
interactions in a superfluid Fermi liquid.

Unless we are dealing with a spin-independent longitudinal field only
fluctuations of the normal component contribute to the polarization. The
corresponding effective vertices should be found from the equations
\cite{Leggett}:%
\begin{equation}
\mathcal{T}_{-}\left(  \mathbf{n}\right)  -\int\frac{d\mathbf{n}^{\prime}%
}{4\pi}\Gamma^{\omega}\left(  \mathbf{nn}^{\prime}\right)  \left[
\kappa\left(  \mathbf{n}^{\prime}\right)  \mathcal{T}_{-}\left(
\mathbf{n}^{\prime}\right)  -\frac{\omega}{\mathbf{qv}^{\prime}}\kappa\left(
\mathbf{n}^{\prime}\right)  \mathcal{T}_{+}\left(  \mathbf{n}^{\prime}\right)
\right]  =\xi_{-}\left(  \mathbf{n}\right)  , \label{psi1n}%
\end{equation}%
\begin{equation}
\mathcal{T}_{+}\left(  \mathbf{n}\right)  +\int\frac{d\mathbf{n}^{\prime}%
}{4\pi}\Gamma^{\omega}\left(  \mathbf{nn}^{\prime}\right)  \left[
\frac{\omega}{\mathbf{qv}^{\prime}}\kappa\left(  \mathbf{n}^{\prime}\right)
\mathcal{T}_{-}\left(  \mathbf{n}^{\prime}\right)  -\left(  \kappa\left(
\mathbf{n}^{\prime}\right)  -2\lambda\left(  \mathbf{n}^{\prime}\right)
\right)  \mathcal{T}_{+}\left(  \mathbf{n}^{\prime}\right)  \right]  =\xi
_{+}\left(  \mathbf{n}\right)  , \label{psi2n}%
\end{equation}
which represent Dyson's equations ideally summing the particle-hole
irreducible diagrams in the ladder approximation. In these equations, the spin
dependence is already taken into account, so $\xi$ is to be taken as a
function of only $\mathbf{p}$, i.e. $\xi\left(  \mathbf{p}\right)  \equiv
\xi\left(  \mathbf{p,}\sigma\right)  =\xi\left(  \mathbf{p,-}\sigma\right)  $
for $S=0$, and $\xi\left(  \mathbf{p}\right)  \equiv\xi\left(  \mathbf{p,}%
\sigma\right)  =-\xi\left(  \mathbf{p,-}\sigma\right)  $ for $S=1$. The $c$
number $\Gamma^{\omega}$ refers to the usual Landau
"quasiparticle-irreducible" scattering amplitude $\Gamma^{\omega}\left(
\mathbf{n,n}^{\prime}\right)  $ as defined in the normal phase; it is to be
taken as the spin-independent or spin-dependent part according to $\xi\left(
\mathbf{p,}\sigma\right)  =\pm\xi\left(  \mathbf{p,-}\sigma\right)  $.

In this way one may calculate the spin, transverse-current, and
helicity-current autocorrelation functions, which are given by the
expressions:
\begin{equation}
K_{\xi}=\rho\int\frac{d\mathbf{n}}{4\pi}\xi_{+}\left(  \mathbf{n}\right)
\left[  \frac{\omega}{\mathbf{qv}}\kappa\left(  \mathbf{n}\right)
\mathcal{T}_{-}\left(  \mathbf{n}\right)  +\left(  \kappa\left(
\mathbf{n}\right)  -2\lambda\left(  \mathbf{n}\right)  \right)  \mathcal{T}%
_{+}\left(  \mathbf{n}\right)  \right]  , \label{Kpn}%
\end{equation}
if $\xi=\xi_{+}$, and%
\begin{equation}
K_{\xi}=\rho\int\frac{d\mathbf{n}}{4\pi}\xi_{-}\left(  \mathbf{n}\right)
\left[  \kappa\left(  \mathbf{n}\right)  \mathcal{T}_{-}\left(  \mathbf{n}%
\right)  -\frac{\omega}{\mathbf{qv}}\kappa\left(  \mathbf{n}\right)
\mathcal{T}_{+}\left(  \mathbf{n}\right)  \right]  \label{Kmn}%
\end{equation}
if $\xi=\xi_{-}$.

We are now in a position to evaluate the autocorrelation functions necessary
for calculating the energy losses from a hot superfluid baryon matter. We
consider the medium response in the vector and axial-vector channels which are
responsible for the neutrino interactions with the medium through neutral weak currents.

\section{Vector channel}

Vector current of a quasiparticle $J^{\mu}$ is a vector in Dirac space
$\left(  \mu=0,1,2,3\right)  $. The corresponding polarization tensor
$\Pi_{\mathrm{V}}^{\mu\nu}\left(  \omega,\mathbf{q}\right)  $ must obey the
current conservation conditions:
\begin{equation}
\Pi_{\mathrm{V}}^{\mu\nu}\left(  \omega,\mathbf{q}\right)  q_{\nu}=0,\ q_{\mu
}\Pi_{\mathrm{V}}^{\mu\nu}\left(  \omega,\mathbf{q}\right)  =0. \label{CC}%
\end{equation}
These equations imply that the polarization tensor can be represented as the
sum of longitudinal (with respect to $\mathbf{q}$) and transverse components%
\begin{equation}
\Pi_{\mathrm{V}}^{\mu\nu}\left(  \omega,\mathbf{q}\right)  =\Pi_{L}\left(
\omega,\mathbf{q}\right)  \left(  1,\frac{\omega}{\mathsf{q}}\mathbf{n}%
_{\mathbf{q}}\right)  ^{\mu}\left(  1,\frac{\omega}{\mathsf{q}}\mathbf{n}%
_{\mathbf{q}}\right)  ^{\nu}+\Pi_{T}\left(  \omega,\mathbf{q}\right)  g^{\mu
i}\left(  \delta^{ij}-n_{\mathbf{q}}^{i}n_{\mathbf{q}}^{j}\right)  g^{j\nu}.
\label{PmnTilde}%
\end{equation}
In this expansion, the longitudinal and transverse polarization functions are
defined as%
\begin{equation}
\Pi_{L}=\Pi^{00},\ \ \ \ \Pi_{T}=\frac{1}{2}\left(  \delta^{ij}-n_{\mathbf{q}%
}^{i}n_{\mathbf{q}}^{j}\right)  \Pi^{ij}. \label{pil}%
\end{equation}
The transverse polarization function can be conveniently evaluated in the
reference frame where the $z$ axis is pointed along the transferred momentum,
so that $\mathbf{n}_{\mathbf{q}}=\left(  0,0,1\right)  $. Then
\begin{equation}
\Pi_{T}\left(  \omega,\mathbf{q}\right)  =\frac{1}{2}\left(  \Pi^{1,1}\left(
\omega,\mathbf{q}\right)  +\Pi^{2,2}\left(  \omega,\mathbf{q}\right)  \right)
. \label{pit}%
\end{equation}

Thus we actually need to calculate only the temporal and transverse components
of the effective vertices.

\subsection{ Longitudinal polarization}

The vector current of a free particle has the nonrelativistic form%
\begin{equation}
j_{V}^{\mu}=\left(  1,\boldsymbol{v}\right)  , \label{jv}%
\end{equation}
where $\boldsymbol{v}=\mathbf{p/}M$ is the particle velocity. In this case we
find
\begin{equation}
\xi^{0}=\xi_{+}^{0}=1,\ \ \ \ \ \xi_{-}^{0}=0, \label{ksi0v}%
\end{equation}%
\begin{equation}
\mathbf{\xi}_{+}^{i}=0,\ \ \ \ \mathbf{\xi}^{i}=\mathbf{\xi}_{-}%
^{i}=\boldsymbol{v}^{i}. \label{ksiiv}%
\end{equation}
Then the longitudinal polarization, $\Pi_{L}=K_{1}\left(  \omega
,\mathbf{q}\right)  ,$can be calculated with the aid of Eqs. (\ref{psi1}%
)-(\ref{Kp}) with $\xi_{+}=1$ and $\xi_{-}=0$.

Before proceeding to the detailed solution of these equations, let us note
that apart from the ground state, Eq. (\ref{psi1}) allows for excitations of
the bound pairs with the orbital momentum $l>0$, if these exist. We will
consider the simplest case of $^{1}S_{0}$ pairing, assuming that the only
possible bound state of the pair corresponds to the zero angular momentum $l$.
This allows us to consider only the zeroth harmonic of the pairing
interaction. In this case the anomalous vertex is independent of the
quasiparticle momentum and the use of the gap equation (\ref{GAPEQ}) allows us
to recast Eq. (\ref{psi1}) as follows:%
\begin{equation}
\mathcal{\tilde{T}}\int\frac{d\mathbf{n}}{4\pi}\left(  \omega^{2}-\left(
\mathbf{qv}\right)  ^{2}\right)  \lambda\left(  \mathbf{n}\right)
=2\Delta\int\frac{d\mathbf{n}}{4\pi}\left(  \omega\lambda\left(
\mathbf{n}\right)  \mathcal{T}_{+}\left(  \mathbf{n}^{\prime}\right)  -\left(
\mathbf{qv}\right)  \lambda\left(  \mathbf{n}\right)  \mathcal{T}_{-}\left(
\mathbf{n}\right)  \right)  . \label{PSI1}%
\end{equation}
Using Eq. (\ref{ksi0v}), we obtain Eqs. (\ref{psi2}) and (\ref{psi3}) in the
form%
\begin{gather}
\mathcal{T}_{-}\left(  \mathbf{n}\right)  +\int\frac{d\mathbf{n}^{\prime}%
}{4\pi}f\left(  \mathbf{nn}^{\prime}\right)  \frac{\mathbf{qv}^{\prime}%
}{\Delta}\lambda\left(  \mathbf{n}^{\prime}\right)  \mathcal{\tilde{T}}\left(
\mathbf{n}^{\prime}\right)  -\int\frac{d\mathbf{n}^{\prime}}{4\pi}f\left(
\mathbf{nn}^{\prime}\right)  \kappa\left(  \mathbf{n}^{\prime}\right)
\mathcal{T}_{-}\left(  \mathbf{n}^{\prime}\right) \nonumber\\
+\int\frac{d\mathbf{n}^{\prime}}{4\pi}f\left(  \mathbf{nn}^{\prime}\right)
\frac{\omega}{\mathbf{qv}^{\prime}}\kappa\left(  \mathbf{n}^{\prime}\right)
\mathcal{T}_{+}\left(  \mathbf{n}^{\prime}\right)  =0, \label{PSI2}%
\end{gather}%
\begin{gather}
\mathcal{T}_{+}\left(  \mathbf{n}\right)  -\frac{\mathrm{\omega}}{\Delta}%
\int\frac{d\mathbf{n}^{\prime}}{4\pi}f\left(  \mathbf{nn}^{\prime}\right)
\lambda\left(  \mathbf{n}^{\prime}\right)  \mathcal{\tilde{T}}\left(
\mathbf{n}^{\prime}\right)  +\int\frac{d\mathbf{n}^{\prime}}{4\pi}f\left(
\mathbf{nn}^{\prime}\right)  \frac{\omega}{\mathbf{qv}^{\prime}}\kappa\left(
\mathbf{n}^{\prime}\right)  \mathcal{T}_{-}\left(  \mathbf{n}^{\prime}\right)
\nonumber\\
-\int\frac{d\mathbf{n}^{\prime}}{4\pi}f\left(  \mathbf{nn}^{\prime}\right)
\left(  \kappa\left(  \mathbf{n}^{\prime}\right)  -2\lambda\left(
\mathbf{n}^{\prime}\right)  \right)  \mathcal{T}_{+}\left(  \mathbf{n}%
^{\prime}\right)  =1. \label{PSI3}%
\end{gather}
The vertex equations can be further simplified in various assumptions about
the amplitude of the particle-hole interaction (\ref{phin}), which can be
expanded in the Legendre polynomials, according to%
\begin{equation}
f\left(  \mathbf{nn}^{\prime}\right)  =\sum_{l}f_{l}P_{l}\left(
\mathbf{nn}^{\prime}\right)  . \label{9}%
\end{equation}
We consider a simplified model with $f_{l}=0$ for $l\geq2$, when the
interaction function is given as%
\begin{equation}
f\left(  \mathbf{nn}^{\prime}\right)  =f_{0}+f_{1}\mathbf{nn}^{\prime}.
\label{ph2}%
\end{equation}

Solution to the set of Eqs. (\ref{PSI1})--(\ref{PSI3}) can be written with the
aid of the following notation:
\begin{align}
\alpha\left(  \omega,\mathrm{q},T\right)   &  \equiv\int\frac{d\mathbf{n}%
}{4\pi}\lambda\left(  \mathbf{n}\right)  ,\ \gamma\left(  \omega
,\mathrm{q},T\right)  \equiv\int\frac{d\mathbf{n}}{4\pi}\lambda\left(
\mathbf{n}\right)  \cos^{2}\theta,\nonumber\\
\eta\left(  \omega,\mathrm{q},T\right)   &  \equiv\int\frac{d\mathbf{n}}{4\pi
}\kappa\left(  \mathbf{n}\right)  ,\ \beta\left(  \omega,\mathrm{q},T\right)
\equiv\int\frac{d\mathbf{n}}{4\pi}\kappa\left(  \mathbf{n}\right)  \cos
^{2}\theta, \label{alphabeta}%
\end{align}%
\[
Q\equiv\eta+\frac{2\alpha\gamma}{s^{2}\alpha-\gamma},\ P\equiv\beta
+\frac{2\gamma^{2}}{s^{2}\alpha-\gamma},
\]
where%
\begin{equation}
s=\frac{\omega}{\mathrm{q}V_{F}}. \label{0}%
\end{equation}
After some algebra, we find:%
\begin{equation}
\mathcal{T}_{+}=\frac{1-f_{1}P}{1-f_{0}\left(  1+f_{1}\left(  s^{2}Q-P\right)
\right)  Q\allowbreak-f_{1}P}, \label{11}%
\end{equation}%
\begin{equation}
\mathcal{T}_{-}=-\frac{sf_{1}Q\cos\theta}{1-f_{0}\left(  1+f_{1}\left(
s^{2}Q-P\right)  \right)  Q\allowbreak-f_{1}P}, \label{12}%
\end{equation}%
\begin{equation}
\mathcal{\tilde{T}}=2\frac{\Delta}{\omega}\frac{s^{2}\left(  \alpha\left(
1-f_{1}\beta\right)  +\gamma\eta f_{1}\right)  }{\left(  s^{2}\alpha
-\gamma\right)  \left[  1-f_{0}\left(  1+f_{1}\left(  s^{2}Q-P\right)
\right)  Q\allowbreak-f_{1}P\right]  }. \label{14}%
\end{equation}
A short calculation of the right-hand side of Eq. (\ref{Kp}) with $\xi_{+}=1$
gives the simple result%
\begin{equation}
\Pi_{L}\left(  \omega,\mathrm{q},T\right)  =\rho\frac{\left(  1+f_{1}\left(
s^{2}Q-P\right)  \right)  Q\allowbreak}{1-f_{0}\left(  1+f_{1}\left(
s^{2}Q-P\right)  \right)  Q\allowbreak-f_{1}P} \label{K1}%
\end{equation}

\subsubsection{BCS limit.}

Notice that the autocorrelation function of the density fluctuations has
already been calculated in various limits. Let us take, for example, the BCS
limit by setting $f_{0}=f_{1}=0$. We then obtain
\begin{equation}
\Pi_{L}^{\mathrm{BCS}}\left(  \omega,\mathrm{q},T\right)  =\rho Q\allowbreak
\equiv\rho\left(  \eta-2\alpha+\frac{2\omega^{2}\alpha^{2}}{\omega^{2}%
\alpha-\mathrm{q}^{2}V_{F}^{2}\gamma}\right)  . \label{15}%
\end{equation}
This expression is in agreement with Eq. (37) of Ref. \cite{L08} if we take
into account the relations $\eta-2\alpha=\Lambda_{00}$,\ $\omega\alpha
=-\Delta\Lambda_{0}$, and\ $\mathrm{q}^{2}V_{F}^{2}\gamma=\Delta q_{i}%
\Lambda_{i}$ connecting our notations and those of Ref. \cite{L08}.

\subsubsection{Limit $\omega,\mathrm{q}V_{F}\ll\Delta$, $T>0$.}

In this limiting case, from Eqs. (\ref{l}), (\ref{k}), and (\ref{alphabeta})
we find (see also Ref. \cite{Leggett})
\begin{equation}
\alpha\simeq\frac{1}{2}+\frac{1}{2}\int\frac{d\mathbf{n}}{4\pi}\int
_{0}^{\infty}d\varepsilon\frac{\left(  \cos^{2}\theta-s^{2}\right)
\varepsilon^{2}/E^{2}}{s^{2}-\left(  \cos^{2}\theta\right)  \varepsilon
^{2}/E^{2}}\frac{dn}{dE}, \label{alpha}%
\end{equation}%
\begin{equation}
\gamma\simeq\frac{1}{6}+\frac{1}{2}\int\frac{d\mathbf{n}}{4\pi}\int
_{0}^{\infty}d\varepsilon\frac{\cos^{2}\theta\ \left(  \cos^{2}\theta
-s^{2}\right)  \varepsilon^{2}/E^{2}}{s^{2}-\left(  \cos^{2}\theta\right)
\varepsilon^{2}/E^{2}}\frac{dn}{dE}, \label{gamma}%
\end{equation}%
\begin{equation}
\eta\simeq\int\frac{d\mathbf{n}}{4\pi}\int_{0}^{\infty}d\varepsilon
\frac{\left(  \cos^{2}\theta\right)  \varepsilon^{2}/E^{2}}{s^{2}-\left(
\cos^{2}\theta\right)  \varepsilon^{2}/E^{2}}\frac{dn}{dE},
\label{etaspacelike}%
\end{equation}%
\begin{equation}
\beta\simeq2\gamma+s^{2}\eta-\frac{1}{3}, \label{beta}%
\end{equation}%
\begin{equation}
P\simeq s^{2}Q-\frac{1}{3}, \label{16}%
\end{equation}
where%
\begin{equation}
\frac{dn}{dE}=\frac{1}{2T}\cosh^{-2}\frac{E}{2T}. \label{17}%
\end{equation}
Then Eq. (\ref{K1}) gives%
\begin{equation}
\Pi_{L}\left(  \omega,\mathrm{q}V_{F}\ll\Delta,T\right)  =\frac{\rho Q\left(
s\right)  }{1-\left[  f_{0}+f_{1}s^{2}/\left(  1+f_{1}/3\right)  \right]
Q\left(  s\right)  }, \label{18}%
\end{equation}
in agreement with the result of Leggett \cite{Leggett}.$\allowbreak$
$\allowbreak$

\subsubsection{Limit $\omega,\mathrm{q}V_{F}\ll\Delta$, $T=0$.}

In the case $T=0$, Eqs. (\ref{alpha})--(\ref{beta}) give%
\begin{equation}
\alpha=\frac{1}{2},\ \gamma=\frac{1}{6},\eta=\beta=0, \label{19}%
\end{equation}
so%
\begin{equation}
Q=\frac{1}{3s^{2}-1},\ P=\frac{1}{3}Q. \label{20}%
\end{equation}
We then obtain
\begin{equation}
\Pi_{L}\left(  \omega,\mathrm{q}V_{F}\ll\Delta,T=0\right)  =\frac{\rho\left(
1+f_{1}/3\right)  \mathrm{q}^{2}V_{F}^{2}/3.}{\omega^{2}-\left(
1+f_{0}\right)  \left(  1+f_{1}/3\right)  \mathrm{q}^{2}V_{F}^{2}/3}.
\label{PL}%
\end{equation}
in agreement with the results obtained in Ref. \cite{Larkin}.

\subsubsection{Time-like momentum transfer, $0<T<T_{c}$}

We are interested in the case of time-like momentum transfer, $\mathrm{q}%
<\omega,$ and $\omega>2\Delta$ taking place in kinematics of the neutrino-pair
emission. Then we deal with the case $\mathrm{q}V_{F}\ll\omega$, i.e.,
$u\equiv s^{-1}\ll1$. In this limit, we have%
\begin{equation}
\operatorname{Re}\gamma=\frac{1}{3}\operatorname{Re}\alpha,\ \beta\sim\eta\sim
u^{2}\alpha, \label{21}%
\end{equation}
Using this fact, we find the functions $Q$ and $P$ in the forms%

\begin{equation}
Q=\eta+2u^{2}\gamma, \label{22}%
\end{equation}%
\begin{equation}
s^{2}Q-P=s^{2}\eta-\beta+2\gamma\label{23}%
\end{equation}

The real and imaginary parts of the functions can be obtained from Eqs.
(\ref{l}), (\ref{k}), and (\ref{alphabeta}). The real part can be evaluated to
the lowest accuracy. We find:%
\begin{equation}
\operatorname{Re}\alpha=-\mathcal{P}\int_{-\infty}^{\infty}\frac{d\varepsilon
}{E}\frac{\Delta^{2}}{\omega^{2}-4E^{2}}\tanh\frac{E}{2T}, \label{Rea}%
\end{equation}%
\begin{equation}
\operatorname{Re}\gamma=-\frac{1}{3}\mathcal{P}\int_{-\infty}^{\infty}%
\frac{d\varepsilon}{E}\frac{\Delta^{2}}{\omega^{2}-4E^{2}}\tanh\frac{E}{2T},
\label{Reg}%
\end{equation}%
\begin{equation}
\operatorname{Re}\eta=\frac{u^{2}}{3}\left(  1+2\mathcal{P}\int_{-\infty
}^{\infty}\frac{d\varepsilon}{E}\frac{\allowbreak\Delta^{2}}{\omega^{2}%
-4E^{2}}\tanh\frac{E}{2T}\right)  , \label{Ree}%
\end{equation}%
\begin{equation}
\operatorname{Re}\beta=\frac{u^{2}}{5}\left(  1+2\mathcal{P}\int_{-\infty
}^{\infty}\frac{d\varepsilon}{E}\frac{\allowbreak\Delta^{2}}{\omega^{2}%
-4E^{2}}\tanh\frac{E}{2T}\right)  , \label{Reb}%
\end{equation}
where the symbol $\mathcal{P}$ means principal value of the integral. In
deriving the last two equalities we used the identity%
\begin{equation}
\frac{1}{2}\int_{-\infty}^{\infty}\frac{d\varepsilon}{E}\frac{\Delta^{2}%
}{E^{2}}\tanh\frac{E}{2T}+\frac{1}{2}\int_{-\infty}^{\infty}d\varepsilon
\frac{\varepsilon^{2}}{E^{2}}\frac{dn}{dE}=1. \label{24}%
\end{equation}
Within a time-like momentum transfer and $\omega>2\Delta$, the imaginary part
of the function arises because of the pole at $\omega=E_{\mathbf{p+q}%
}+E_{\mathbf{p}}$. We calculate the imaginary contributions up to the higher
accuracy and find
\begin{align}
\operatorname{Im}\alpha &  =\pi\frac{\Delta^{2}}{\omega\sqrt{\omega
^{2}-4\Delta^{2}}}\Theta\left(  \omega-2\Delta\right)  \tanh\frac{\omega}%
{4T}\nonumber\\
&  \times\left(  \allowbreak1+\frac{1}{3}u^{2}\left(  \frac{\omega^{2}%
+4\Delta^{2}}{\omega^{2}-4\Delta^{2}}\allowbreak-\frac{\omega^{2}-4\Delta^{2}%
}{16T^{2}}\cosh^{-2}\frac{\omega}{4T}\right)  \allowbreak\right)  ,
\label{Ima}%
\end{align}%
\begin{align}
\operatorname{Im}\gamma &  =\frac{\pi}{3}\frac{\Delta^{2}}{\omega\sqrt
{\omega^{2}-4\Delta^{2}}}\Theta\left(  \omega-2\Delta\right)  \tanh
\frac{\omega}{4T}\nonumber\\
&  \times\left(  1\allowbreak+\frac{3}{5}u^{2}\left(  \frac{8\Delta^{2}%
+\omega^{2}}{\omega^{2}-4\Delta^{2}}-\frac{\omega^{2}-4\Delta^{2}}{16T^{2}%
}\cosh^{-2}\frac{\omega}{4T}\allowbreak\right)  \allowbreak\allowbreak\right)
\label{Img}%
\end{align}%
\begin{align}
\operatorname{Im}\eta &  =-\frac{2\pi}{3}\frac{u^{2}\Delta^{2}}{\omega
\sqrt{\omega^{2}-4\Delta^{2}}}\Theta\left(  \omega-2\Delta\right)  \tanh
\frac{\omega}{4T}\nonumber\\
&  \times\left(  1+\frac{6}{5}u^{2}\left(  \frac{\omega^{2}+2\Delta^{2}%
}{\omega^{2}-4\Delta^{2}}-\frac{\omega^{2}-4\Delta^{2}}{32T^{2}}\cosh
^{-2}\frac{\omega}{4T}\right)  \right)  \allowbreak, \label{Ime}%
\end{align}%
\begin{align}
\operatorname{Im}\beta &  =-\frac{2\pi}{5}\frac{u^{2}\Delta^{2}}{\omega
\sqrt{\omega^{2}-4\Delta^{2}}}\Theta\left(  \omega-2\Delta\right)  \tanh
\frac{\omega}{4T}\nonumber\\
&  \times\left(  \allowbreak1+\frac{25}{28}u^{2}\left(  \allowbreak
\frac{\omega^{2}+4\Delta^{2}}{\omega^{2}-4\Delta^{2}}\allowbreak
-\allowbreak\frac{\omega^{2}-4\Delta^{2}}{80T^{2}}\cosh^{-2}\frac{\omega}%
{4T}\allowbreak\allowbreak\allowbreak\allowbreak\right)  \right)  ,
\label{Imb}%
\end{align}
where $\Theta\left(  x\right)  $ is the ordinary Heaviside step function.

We also find:%
\begin{equation}
Q=\frac{u^{2}}{3}-i\frac{2\pi}{5}\frac{u^{4}\Delta^{2}\Theta\left(
\omega-2\Delta\right)  }{\omega\sqrt{\omega^{2}-4\Delta^{2}}}\tanh\frac
{\omega}{4T}, \label{25}%
\end{equation}%
\begin{align}
s^{2}Q-P  &  =\frac{1}{3}+i\frac{5\pi}{14}\frac{u^{4}\Delta^{2}\Theta\left(
\omega-2\Delta\right)  }{\omega\left(  \omega^{2}-4\Delta^{2}\right)
\sqrt{\omega^{2}-4\Delta^{2}}}\tanh\frac{\omega}{4T}\nonumber\\
&  \times\left(  \allowbreak\frac{\omega^{2}+4\Delta^{2}}{\omega^{2}%
-4\Delta^{2}}\allowbreak-\allowbreak\frac{\omega^{2}-4\Delta^{2}}{80T^{2}%
}\cosh^{-2}\frac{\omega}{4T}\allowbreak\allowbreak\allowbreak\allowbreak
\right)  , \label{26}%
\end{align}
and%
\begin{align}
P  &  =\frac{u^{2}}{5}\left(  1+\frac{8}{9}\mathcal{P}\int_{-\infty}^{\infty
}\frac{d\varepsilon}{E}\frac{\allowbreak\Delta^{2}}{\omega^{2}-4E^{2}}%
\tanh\frac{E}{2T}\right) \nonumber\\
&  -i\frac{2\pi}{5}\frac{u^{2}\Delta^{2}\Theta\left(  \omega-2\Delta\right)
}{\omega\sqrt{\omega^{2}-4\Delta^{2}}}\tanh\frac{\omega}{4T}. \label{27}%
\end{align}

Having these formulas at hand, we can evaluate the real and imaginary parts of
the longitudinal polarization function (\ref{K1}). After a little algebra, we
obtain%
\begin{align}
\Pi_{L}\left(  \omega,\mathrm{q},T\right)   &  =\rho\frac{1}{3}V_{F}%
^{2}\left(  1+\frac{1}{3}f_{1}\right)  \frac{\mathrm{q}^{2}}{\omega^{2}%
}\nonumber\\
&  -i\frac{2\pi}{5}\rho V_{F}^{4}\left(  1+\frac{1}{3}f_{1}\right)  ^{2}%
\frac{\mathrm{q}^{4}\Delta^{2}\Theta\left(  \omega-2\Delta\right)  }%
{\omega^{5}\sqrt{\omega^{2}-4\Delta^{2}}}\tanh\frac{\omega}{4T}, \label{L}%
\end{align}
As one can see from this expression the spherical harmonic of the pairing
interaction does not affect the longitudinal polarization in the
high-frequency limit $\omega\gg\mathrm{q}V_{F}$. If we set $f_{1}=0$, this
expression reproduces the result of the BCS approximation [see Eq. (48) in
Ref. \cite{L08}].

\subsection{Transverse polarization}

As explained above, the transverse field does not affect the anomalous
self-energy of a quasiparticle. Therefore the transverse-current
autocorrelation function
\begin{equation}
K_{T}\left(  \omega,\mathrm{q}\right)  =\frac{1}{2}\left(  K_{\xi_{-}=v_{1}%
}\left(  \omega,\mathrm{q}\right)  +K_{\xi_{-}=v_{2}}\left(  \omega
,\mathrm{q}\right)  \right)  .\label{28}%
\end{equation}
can be evaluated with the aid of Eqs. (\ref{psi1n}), (\ref{psi2n}), and
(\ref{Kmn}) with $\xi_{+}=0$ and $\xi_{-}^{i}=v_{\perp}^{i}$, where
$\boldsymbol{v}_{\perp}=\left(  v\sin\theta\cos\varphi,v\sin\theta\sin
\varphi,0\right)  $. The particle-hole interaction (\ref{ph2}) can be written
as
\begin{equation}
f_{0}+f_{1}\mathbf{nn}^{\prime}\equiv f_{0}+f_{1}\left(  \cos\theta\cos
\theta^{\prime}+\sin\theta\sin\theta^{\prime}\cos\left(  \varphi
-\varphi^{\prime}\right)  \right)  .\label{29}%
\end{equation}
The sets of equations for different $i=\left(  1,2\right)  $ are decoupled,
and we find:%
\begin{equation}
\mathcal{T}_{+}^{\left(  i\right)  }\left(  \mathbf{n}\right)
=0,\ \mathcal{T}_{-}^{\left(  i\right)  }=\frac{v_{\perp}^{\left(  i\right)
}}{1+f_{1}\left(  \eta-\beta\right)  /2}\label{30}%
\end{equation}
and%
\begin{equation}
K_{T}\left(  \omega,\mathrm{q}\right)  =\frac{\rho}{2}\frac{V_{F}^{2}\left(
\eta-\beta\right)  }{1+f_{1}\left(  \eta-\beta\right)  /2}\label{Kperp}%
\end{equation}

In the case $\mathrm{q}<\omega,$ and $\omega>2\Delta$, using Eqs. (\ref{Ree}),
(\ref{Reb}), (\ref{Ime}), and (\ref{Imb}), we find%
\begin{align}
\eta-\beta &  =\frac{2}{15}u^{2}\left(  1+2\mathcal{P}\int_{-\infty}^{\infty
}\frac{d\varepsilon}{E}\frac{\allowbreak\Delta^{2}}{\omega^{2}-4E^{2}}%
\tanh\frac{E}{2T}\right) \nonumber\\
&  -i\frac{4\pi}{15}\frac{u^{2}\Delta^{2}}{\omega\sqrt{\omega^{2}-4\Delta^{2}%
}}\tanh\frac{\omega}{4T}. \label{31}%
\end{align}
Up to accuracy $V_{F}^{4}$ from Eq. (\ref{Kperp}), we obtain%
\begin{align}
K_{T}\left(  \omega,\mathrm{q}\right)   &  =\frac{1}{15}\rho V_{F}^{4}%
\frac{\mathrm{q}^{2}}{\omega^{2}}\left(  1+2\mathcal{P}\int_{-\infty}^{\infty
}\frac{d\varepsilon}{E}\frac{\allowbreak\Delta^{2}}{\omega^{2}-4E^{2}}%
\tanh\frac{E}{2T}\right) \nonumber\\
&  -i\frac{2\pi}{15}\rho V_{F}^{4}\frac{\mathrm{q}^{2}\Delta^{2}}{\omega
^{3}\sqrt{\omega^{2}-4\Delta^{2}}}\tanh\frac{\omega}{4T}. \label{T}%
\end{align}
This expression coincides with that of the BCS approximation \cite{L08}. We
see that in the high-frequency limit, $u\ll1,$ the first two harmonics of the
particle-hole interaction do not affect the transverse polarization of the medium.

\section{Axial channel}

Since only the normal component contributes to the spin fluctuations, the
axial effective vertices should be found from Eqs. (\ref{psi1n}) and
(\ref{psi2n}), and the corresponding correlation functions are given by Eqs.
(\ref{Kpn}) and (\ref{Kmn}). We now focus on this calculation.

The operator of the axial-vector current is a Dirac pseudovector. For a free
particle, it is of the nonrelativistic form $\left(  \mu=0,1,2,3\right)  $%
\begin{equation}
\hat{\jmath}_{A}^{\mu}=\left(
{\textstyle\sum_{i}}
\hat{\sigma}_{i}v_{i}\mathbf{,}\hat{\sigma}_{1}\mathbf{,}\hat{\sigma}%
_{2}\mathbf{,}\hat{\sigma}_{3}\right)  , \label{32}%
\end{equation}
where $\boldsymbol{v}=\mathbf{p}/M$ is the particle velocity, and $\hat
{\sigma}_{i}$ are Pauli spin matrices. For $S=1$, the exchange part of the
particle-hole interaction is to be taken as
\begin{equation}
g\left(  \mathbf{nn}^{\prime}\right)
{\textstyle\sum_{i}}
\hat{\sigma}_{i}\hat{\sigma}_{i}^{\prime}=\frac{1}{4}g\left(  \mathbf{nn}%
^{\prime}\right)  , \label{33}%
\end{equation}
and
\begin{equation}
\xi_{+}^{\mu}=v\delta_{\mu0},\ \xi_{-}^{\mu}=\delta_{\mu,i}. \label{34}%
\end{equation}
Then for a space part of the correlation tensor ($i,j=1,2,3$) we find
$K_{A}^{ij}=\delta_{ij}K_{A}$, where
\begin{equation}
K_{A}\left(  \omega,\mathrm{q}\right)  =\rho\int\frac{d\mathbf{n}}{4\pi
}\left[  \kappa\left(  \mathbf{n}\right)  \mathcal{T}_{-}\left(
\mathbf{n}\right)  -\frac{\omega}{\mathbf{qv}}\kappa\left(  \mathbf{n}\right)
\mathcal{T}_{+}\left(  \mathbf{n}\right)  \right]  , \label{KIJ}%
\end{equation}
and the full vertices are to satisfy the equations%
\begin{equation}
\mathcal{T}_{-}\left(  \mathbf{n}\right)  -\frac{1}{4}\int\frac{d\mathbf{n}%
^{\prime}}{4\pi}g\left(  \mathbf{nn}^{\prime}\right)  \left[  \kappa\left(
\mathbf{n}^{\prime}\right)  \mathcal{T}_{-}\left(  \mathbf{n}^{\prime}\right)
-\frac{\omega}{\mathbf{qv}^{\prime}}\kappa\left(  \mathbf{n}^{\prime}\right)
\mathcal{T}_{+}\left(  \mathbf{n}^{\prime}\right)  \right]  =1, \label{35}%
\end{equation}%
\begin{equation}
\mathcal{T}_{+}\left(  \mathbf{n}\right)  +\frac{1}{4}\int\frac{d\mathbf{n}%
^{\prime}}{4\pi}g\left(  \mathbf{nn}^{\prime}\right)  \left[  \frac{\omega
}{\mathbf{qv}^{\prime}}\kappa\left(  \mathbf{n}^{\prime}\right)
\mathcal{T}_{-}\left(  \mathbf{n}^{\prime}\right)  -\left(  \kappa\left(
\mathbf{n}^{\prime}\right)  -2\lambda\left(  \mathbf{n}^{\prime}\right)
\right)  \mathcal{T}_{+}\left(  \mathbf{n}^{\prime}\right)  \right]  =0.
\label{36}%
\end{equation}
The temporal component is of the form:
\begin{equation}
K_{A}^{00}\left(  \omega,\mathrm{q}\right)  =\rho v\int\frac{d\mathbf{n}}%
{4\pi}\left[  \frac{\omega}{\mathbf{qv}}\kappa\left(  \mathbf{n}\right)
\mathcal{T}_{-}^{0}\left(  \mathbf{n}\right)  +\left(  \kappa\left(
\mathbf{n}\right)  -2\lambda\left(  \mathbf{n}\right)  \right)  \mathcal{T}%
_{+}^{0}\left(  \mathbf{n}\right)  \right]  , \label{K0m}%
\end{equation}
where the full vertices should be found from the following set of equations%
\begin{equation}
\mathcal{T}_{-}^{0}\left(  \mathbf{n}\right)  -\frac{1}{4}\int\frac
{d\mathbf{n}^{\prime}}{4\pi}g\left(  \mathbf{nn}^{\prime}\right)  \left[
\kappa\left(  \mathbf{n}^{\prime}\right)  \mathcal{T}_{-}^{0}\left(
\mathbf{n}^{\prime}\right)  -\frac{\omega}{\mathbf{qv}^{\prime}}\kappa\left(
\mathbf{n}^{\prime}\right)  \mathcal{T}_{+}^{0}\left(  \mathbf{n}^{\prime
}\right)  \right]  =0, \label{37}%
\end{equation}%
\begin{equation}
\mathcal{T}_{+}^{0}\left(  \mathbf{n}\right)  +\frac{1}{4}\int\frac
{d\mathbf{n}^{\prime}}{4\pi}g\left(  \mathbf{nn}^{\prime}\right)  \left[
\frac{\omega}{\mathbf{qv}^{\prime}}\kappa\left(  \mathbf{n}^{\prime}\right)
\mathcal{T}_{-}^{0}\left(  \mathbf{n}^{\prime}\right)  -\left(  \kappa\left(
\mathbf{n}^{\prime}\right)  -2\lambda\left(  \mathbf{n}^{\prime}\right)
\right)  \mathcal{T}_{+}^{0}\left(  \mathbf{n}^{\prime}\right)  \right]  =v.
\label{38}%
\end{equation}
Mixed space-time components are given by%
\begin{equation}
K_{A}^{0i}\left(  \omega,\mathrm{q}\right)  =\rho v\int\frac{d\mathbf{n}}%
{4\pi}\left[  \frac{\omega}{\mathbf{qv}}\kappa\left(  \mathbf{n}\right)
\mathcal{T}_{-}\left(  \mathbf{n}\right)  +\left(  \kappa\left(
\mathbf{n}\right)  -2\lambda\left(  \mathbf{n}\right)  \right)  \mathcal{T}%
_{+}\left(  \mathbf{n}\right)  \right]  , \label{39}%
\end{equation}%
\begin{equation}
K_{A}^{i0}\left(  \omega,\mathrm{q}\right)  =\rho\int\frac{d\mathbf{n}}{4\pi
}\left[  \kappa\left(  \mathbf{n}\right)  \mathcal{T}_{-}^{0}\left(
\mathbf{n}\right)  -\frac{\omega}{\mathbf{qv}}\kappa\left(  \mathbf{n}\right)
\mathcal{T}_{+}^{0}\left(  \mathbf{n}\right)  \right]  . \label{Ki0}%
\end{equation}

To obtain a solution in reasonably simple form, we approximate the interaction
amplitude by its first two harmonics, according $g\left(  \mathbf{nn}^{\prime
}\right)  \equiv g_{0}+g_{1}\mathbf{nn}^{\prime}$. Then we find the full
vertices in the form%
\begin{align}
\mathcal{T}_{+}^{0}  &  =\frac{v}{1+g_{0}\left(  2\alpha-\eta\left(
1+B_{1}s^{2}\eta\right)  \right)  /4},\nonumber\\
\mathcal{T}_{-}^{0}\left(  \mathbf{n}\right)   &  =-\frac{vB_{1}s\eta
\cos\theta}{1+g_{0}\left(  2\alpha-\eta\left(  1+B_{1}s^{2}\eta\right)
\right)  /4}, \label{Tp}%
\end{align}%
\begin{align}
\mathcal{T}_{+}\left(  \mathbf{n}\right)   &  =-\frac{B_{2}\eta s\cos\theta
}{1-g_{0}\eta\left(  1+B_{2}s^{2}\eta\right)  /4},\nonumber\\
\mathcal{T}_{-}  &  =\frac{1}{1-g_{0}\eta\left(  1+B_{2}s^{2}\eta\right)  /4},
\label{Tm}%
\end{align}
where
\begin{equation}
B_{1}\left(  \omega,\mathrm{q}\right)  \equiv\frac{1}{4}g_{1}\left(
1-\frac{1}{4}g_{1}\beta\right)  ^{-1}, \label{40}%
\end{equation}%
\begin{equation}
B_{2}\left(  \omega,\mathrm{q}\right)  \equiv\frac{1}{4}g_{1}\left(
1-\frac{1}{4}g_{1}\left(  \beta-2\gamma\right)  \right)  ^{-1}. \label{41}%
\end{equation}
Simple algebraic calculations yield the following autocorrelation functions:%
\begin{equation}
K_{A}^{00}\left(  \omega,\mathrm{q}\right)  =-\rho v^{2}\frac{2\alpha
-\eta\left(  1-B_{1}s^{2}\eta\right)  }{1+g_{0}\left(  2\alpha-\eta\left(
1+B_{1}s^{2}\eta\right)  \right)  /4}, \label{K00}%
\end{equation}
and%
\begin{equation}
K_{A}^{ij}\left(  \omega,\mathrm{q}\right)  =\delta_{ij}\rho\frac{\eta\left(
1+B_{2}s^{2}\eta\right)  }{1-g_{0}\eta\left(  1+B_{2}s^{2}\eta\right)  /4},
\label{Kij}%
\end{equation}
Mixed components $K_{A}^{0i}$ and $K_{A}^{i0}$ are given by the integrals
(\ref{K0m}) and (\ref{Ki0}), where, according to Eqs. (\ref{parity}),
(\ref{Tp}) and (\ref{Tm}), the integrands are odd in $\cos\theta$. Therefore
the mixed polarization vanishes:
\begin{equation}
K_{A}^{0i}\left(  \omega,\mathrm{q}\right)  =K_{A}^{i0}\left(  \omega
,\mathrm{q}\right)  =0. \label{K0j}%
\end{equation}

Let us consider various limits in the expressions obtained above. For
arbitrary temperature $T>0$ and $\omega,\mathrm{q}V_{F}\ll\Delta$, according
to Eq. (\ref{beta}), we have
\begin{equation}
\beta=2\gamma+s^{2}\eta-\frac{1}{3}, \label{42}%
\end{equation}
and%
\begin{equation}
B_{2}=\frac{1}{4}g_{1}\frac{1}{1-g_{1}\left(  s^{2}\eta-\frac{1}{3}\right)
/4}. \label{43}%
\end{equation}
Then the spin-density autocorrelation function (\ref{Kij}) reproduces the
result obtained in Ref. \cite{Leggett}%
\begin{equation}
K_{A}^{ij}\left(  \omega,\mathrm{q}\right)  =\delta_{ij}\rho\frac{\eta\left(
s\right)  }{1-\eta\left(  s\right)  J\left(  s\right)  }, \label{Kijspacelike}%
\end{equation}
where%
\begin{equation}
J\left(  s\right)  =\frac{1}{4}\left(  g_{0}+\frac{s^{2}g_{1}}{1+g_{1}%
/12}\right)  , \label{44}%
\end{equation}
and $\eta\left(  s\right)  $ is given by Eq. (\ref{etaspacelike}).

Next we consider the case of time-like momentum transfer when $\mathrm{q}%
V_{F}\ll\Delta$, $\omega>2\Delta$, and thus $u\equiv s^{-1}\ll1$. From Eqs.
(\ref{Rea}), (\ref{Reg}) and (\ref{Ima}), (\ref{Img}), we find in this limit
\begin{equation}
\gamma\left(  \omega,T\right)  \simeq\frac{1}{3}\alpha\left(  \omega,T\right)
. \label{gam13}%
\end{equation}
For $\omega>0$, we obtain%
\begin{equation}
\operatorname{Im}K_{A}^{00}\left(  \omega,\mathrm{q}\right)  \simeq-2\pi\rho
v^{2}\frac{\Delta^{2}}{\omega\sqrt{\omega^{2}-4\Delta^{2}}}\frac{\Theta\left(
\omega-2\Delta\right)  }{\left\vert 1+g_{0}\alpha\left(  \omega,T\right)
/2\right\vert ^{2}}\tanh\frac{\omega}{4T}, \label{ImK00}%
\end{equation}
and%
\begin{equation}
\operatorname{Im}K_{A}^{ij}\left(  \omega,\mathrm{q}\right)  =-\delta
_{ij}\frac{2\pi}{3}\rho V_{F}^{2}\frac{\mathrm{q}^{2}\Delta^{2}\Theta\left(
\omega-2\Delta\right)  }{\omega^{3}\sqrt{\omega^{2}-4\Delta^{2}}}\frac{\left(
1+g_{1}/12\right)  ^{2}}{\left\vert 1+g_{1}\alpha\left(  \omega,T\right)
/6\right\vert ^{2}}\tanh\frac{\omega}{4T}, \label{ImKij}%
\end{equation}
where%
\begin{equation}
\alpha\left(  \omega,T\right)  =-\mathcal{P}\int_{-\infty}^{\infty}%
\frac{d\varepsilon}{E}\frac{\Delta^{2}}{\omega^{2}-4E^{2}}\tanh\frac{E}%
{2T}+i\pi\frac{\Delta^{2}\Theta\left(  \omega-2\Delta\right)  }{\omega
\sqrt{\omega^{2}-4\Delta^{2}}}\tanh\frac{\omega}{4T}. \label{45}%
\end{equation}

\section{Neutrino energy losses caused by pair recombination}

As an application of the obtained results we consider the neutrino-pair
emission through neutral weak currents occurring at the recombination of
quasiparticles into the $^{1}S_{0}$ condensate. The process is kinematically
allowed thanks to the existence of a superfluid energy gap $\Delta$, which
admits the quasiparticle transitions with time-like momentum transfer
$q=\left(  \omega,\mathbf{q}\right)  $, as required by the final neutrino pair.

We consider the total energy which is emitted into neutrino pairs per unit
volume and time which is given by the following formula (see details, e.g., in
Ref. \cite{L01}):
\begin{equation}
\epsilon=-\left(  \frac{G_{F}}{2\sqrt{2}}\right)  ^{2}\sum_{\nu}\int
\omega\;\frac{2\ \mathrm{Im}\Pi_{\mathrm{weak}}^{\mu\nu}\left(  q\right)
\ \mathrm{Tr}\left(  l_{\mu}l_{\nu}^{\ast}\right)  }{\exp\left(  \frac{\omega
}{T}\right)  -1}\frac{d^{3}q_{1}}{2\omega_{1}(2\pi)^{3}}\frac{d^{3}q_{2}%
}{2\omega_{2}(2\pi)^{3}}, \label{Qtot}%
\end{equation}
where $G_{F}$ is the Fermi coupling constant, $l_{\mu}$ is the neutrino weak
current, and $\Pi_{\mathrm{weak}}^{\mu\nu}$ is the retarded weak polarization
tensor of the medium. The integration goes over the phase volume of neutrinos
and antineutrinos of total energy $\omega=\omega_{1}+\omega_{2}$ and total
momentum $\mathbf{q=q}_{1}+\mathbf{q}_{2}$. The symbol $\sum_{\nu}%
$\ \ indicates that summation over the three neutrino types has to be performed.

By inserting $\int d^{4}q\delta^{\left(  4\right)  }\left(  q-q_{1}%
-q_{2}\right)  =1$ in this equation, and making use of the Lenard's integral
\
\begin{equation}
\int\frac{d^{3}q_{1}}{2\omega_{1}}\frac{d^{3}q_{2}}{2\omega_{2}}%
\delta^{\left(  4\right)  }\left(  q-q_{1}-q_{2}\right)  \mathrm{Tr}\left(
l_{\mu}l_{\nu}^{\ast}\right)  =\frac{4\pi}{3}\left(  q_{\mu}q_{\nu}%
-q^{2}g_{\mu\nu}\right)  \Theta\left(  q^{2}\right)  \Theta\left(
\omega\right)  ,\label{47}%
\end{equation}
where $g_{\mu\nu}=\mathsf{\mathrm{diag}}(1,-1,-1,-1)$ is the signature tensor,
we can write%
\begin{equation}
\epsilon=-\frac{G_{F}^{2}\mathcal{N}_{\nu}}{48\pi^{4}}\int_{0}^{\infty}%
d\omega\int_{0}^{\omega}d\mathrm{q}\;\mathrm{q}^{2}\frac{\omega}{\exp\left(
\frac{\omega}{T}\right)  -1}\mathrm{Im}\Pi_{\mathrm{weak}}^{\mu\nu}\left(
q\right)  \left(  q_{\mu}q_{\nu}-q^{2}g_{\mu\nu}\right)  ,\label{QQQ}%
\end{equation}
where $\mathcal{N}_{\nu}=3$ is the number of neutrino flavors.

In general, the weak polarization tensor of the medium is a sum of the
vector-vector, axial-axial, and mixed terms. However, the medium polarization
in the vector channel can be neglected, because the imaginary part of the
longitudinal and transverse polarization functions is proportional to
$V_{F}^{4}\lll1$, as given by Eqs. (\ref{L}) and (\ref{T}). (See also Refs.
\cite{LP06}, \cite{L08} for details). The mixed axial-vector polarization has
to be an antisymmetric tensor, and its contraction in Eq. (\ref{QQQ}) with the
symmetric tensor $q_{\mu}q_{\nu}-q^{2}g_{\mu\nu}$ vanishes. Thus only
polarization in the axial channel should be taken into account.

We then obtain $\mathrm{Im}\Pi_{\mathrm{weak}}^{\mu\nu}\simeq C_{A}%
^{2}\mathrm{Im}K_{A}^{\mu\nu}$, where $C_{A}$ is the axial weak coupling
constant of the baryon. Making use of Eqs. (\ref{K0j}), (\ref{ImK00}), and
(\ref{ImKij}), we find%
\begin{align}
&  \mathrm{Im}\Pi^{\mu\nu}\left(  q\right)  \left(  q_{\mu}q_{\nu}-q^{2}%
g_{\mu\nu}\right)  =-\frac{2}{\pi}C_{A}^{2}p_{F}M^{\ast}V_{F}^{2}\frac
{\Delta^{2}\Theta\left(  \omega-2\Delta\right)  }{\omega\sqrt{\omega
^{2}-4\Delta^{2}}}\tanh\frac{\omega}{4T}\nonumber\\
&  \times\mathrm{q}^{2}\left(  \frac{M^{\ast2}}{M^{2}}\frac{\left(
1+g_{1}/12\right)  ^{2}}{\left\vert 1+g_{0}\alpha\left(  \omega,T\right)
/2\right\vert ^{2}}+\frac{1}{\left\vert 1+g_{1}\alpha\left(  \omega,T\right)
/6\right\vert ^{2}}\left(  1-\frac{2}{3}\frac{\mathrm{q}^{2}}{\omega^{2}%
}\right)  \right)  . \label{IPA}%
\end{align}
By inserting this into Eq. (\ref{QQQ}) and performing integration over
$d\mathrm{q}$, we obtain the neutrino emissivity in the axial channel, which
can be represented in the form%
\begin{align}
\epsilon &  =\frac{4}{15\pi^{5}}G_{F}^{2}C_{A}^{2}\mathcal{N}_{\nu}%
p_{F}M^{\ast}V_{F}^{2}T^{7}y^{2}\int_{0}^{\infty}dx\frac{z^{4}}{\left(
e^{z}+1\right)  ^{2}}\nonumber\\
&  \times\left(  \frac{M^{\ast2}}{M^{2}}\frac{\left(  1+g_{1}/12\right)  ^{2}%
}{\left\vert 1+g_{0}\alpha\left(  y,z\right)  /2\right\vert ^{2}}+\frac
{11}{21}\frac{1}{\left\vert 1+g_{1}\alpha\left(  y,z\right)  /6\right\vert
^{2}}\right)  , \label{EA}%
\end{align}
where $y=\Delta/T$ and $z=\sqrt{x^{2}+y^{2}}$. The function $\alpha\left(
y,z\right)  $ is given by
\begin{align}
\alpha\left(  y,z\right)   &  =-\frac{1}{2}\mathcal{P}\int_{y}^{\infty}%
\frac{d\upsilon}{\sqrt{\upsilon^{2}-y^{2}}}\frac{y^{2}}{\left(  z^{2}%
-\upsilon^{2}\right)  }\tanh\frac{\upsilon}{2}\nonumber\\
&  +i\frac{\pi}{4}\frac{y^{2}}{z\sqrt{z^{2}-y^{2}}}\tanh\frac{z}{2},
\label{48}%
\end{align}

Some comments on the approximations done in previous works would be here
appropriate. In Refs. \cite{FRS76}, \cite{LP06}, \cite{L08}, \cite{Kaminker},
\cite{Jaikumar} the calculation of the neutrino emissivity is performed in the
BCS approximation, i.e. the authors discard Fermi-liquid interactions in a
superfluid system. The attempt to take into account the particle-hole
interactions was undertaken recently in Ref. \cite{KV08}. However, though the
authors state the important role of the particle-hole interactions, their
final result for neutrino emissivity contains no Landau parameters
characterizing this interaction [see Eq.(35) of Ref. \cite{KV08}]. As a matter
of fact this means that the Fermi-liquid effects have been discarded in this
calculation and the result also corresponds to the BCS approximation.

Thus only the BCS limit of our Eq. (\ref{EA}) can be compared with the
previous calculations. Setting $g_{0}=g_{1}=0$, we obtain%
\begin{equation}
\epsilon^{\mathrm{BCS}}=\frac{4}{15\pi^{5}}\left(  \frac{M^{\ast2}}{M^{2}%
}+\frac{11}{21}\right)  G_{F}^{2}C_{A}^{2}\mathcal{N}_{\nu}p_{F}M^{\ast}%
V_{F}^{2}T^{7}y^{2}\int_{0}^{\infty}dx\frac{z^{4}}{\left(  e^{z}+1\right)
^{2}}, \label{EABCS}%
\end{equation}
where $y=\Delta/T$ and $z=\sqrt{x^{2}+y^{2}}$.

Although this expression reproduces the known BCS result for the neutrino
emissivity in the axial channel, we recall that the total neutrino emissivity,
as given by this formula, is suppressed as $V_{F}^{2}$ with respect to the
earlier results because the vector channel is practically closed. Second term
in the brackets was for the first time obtained in Ref. \cite{FRS76}. The
first term is the same as in Ref. \cite{Kaminker}. Notice that this term
originating from the temporal component of the axial-vector current is lost in
Ref. \cite{SR}.

We also do not support the result obtained in Ref. \cite{KV08}, where one more
term is suggested due to the mixed space-temporal polarization of the medium.
In our calculations, the mixed contribution, being odd in $\cos\theta$,
vanishes on angle integration: see our Eq. (\ref{K0j}). This agrees with the
results obtained in Refs. \cite{FRS76}, \cite{Kaminker}, \cite{Jaikumar}.

The temperature dependence of the energy losses, as obtained in Refs.
\cite{KV08}, \cite{SR}, also is not convincing, because the imaginary parts of
the polarization functions are calculated for zero temperature when no broken
Cooper pair exists. The temperature dependence, as given in our Eq.
(\ref{EABCS}), has been repeatedly obtained by many authors before (see, e.g.,
Refs. \cite{FRS76}, \cite{Kaminker}, \cite{Jaikumar}). This dependence follows
directly from the kinematics of the reaction and statistics of the
pair-correlated fermions.

According to our Eq. (\ref{IPA}), the imaginary part of the retarded
polarization tensor substantially depends on the temperature. This dependence
may be easily understood in the BCS approximation. In this case,
\[
\mathrm{Im}\Pi^{\mu\nu}\propto\tanh\frac{\omega}{4T},
\]
and (besides the temperature dependence of the energy gap) the
temperature-dependent factor in the integrand of Eq. (\ref{QQQ}),%

\begin{equation}
\frac{1}{\exp\frac{\omega}{T}-1}\tanh\frac{\omega}{4T}\equiv\frac{1}{\left(
\exp\frac{\omega}{2T}+1\right)  ^{2}}, \label{df}%
\end{equation}
represents the product of occupation numbers in the initial state of two
recombining quasiparticles. Indeed, the dominant contribution to the phase
integral enters from the quasiparticle momenta near the Fermi surface. As the
neutrino-pair momentum $\mathrm{q}\sim T_{c}\ll\mathrm{p}_{F}$, one can
neglect $\mathbf{q}$ in the momentum conservation $\delta$ function, thus
obtaining $\mathbf{p}^{\prime}=-\mathbf{p}$. After this simplification, the
energies of initial quasiparticles are $E_{\mathbf{p}^{\prime}}=E_{\mathbf{p}%
}=\omega/2$.

\section{Numerical evaluation}

In Eq. (\ref{EA}), the temperature dependence of the emissivity enters by
means of parameter
\begin{equation}
y=\frac{\Delta\left(  T\right)  }{T}=\frac{\Delta\left(  0\right)  }{T_{c}%
}\frac{\Delta\left(  \tau\right)  }{\tau\Delta\left(  0\right)  } \label{y}%
\end{equation}
with $\tau=T/T_{c}$, where T$_{c}$ is the superfluid transition temperature.
For a singlet-state pairing $\Delta\left(  0\right)  /T_{c}=1.\,\allowbreak76$
(see, e.g., Ref. \cite{AGD}), therefore the function $y$ depends on the
dimensionless temperature $\tau$ only. Thus, the emissivity in Eq. (\ref{EA}),
in the standard physical units, can be written as
\begin{align}
\epsilon &  =\frac{4G_{F}^{2}p_{F}M^{\ast}}{15\pi^{5}\hbar^{10}c^{6}}\left(
k_{B}T\right)  ^{7}\mathcal{N}_{\nu}C_{A}^{2}V_{F}^{2}\left(  \frac
{\Delta\left(  0\right)  }{T_{c}}\right)  ^{2}F\left(  \tau\right) \nonumber\\
&  =1.\,\allowbreak17\times10^{21}\mathcal{N}_{\nu}\left(  \frac{M^{\ast}%
}{M_{p}}\right)  ^{2}\left(  \frac{V_{F}}{c}\right)  ^{3}\left(  \frac{T_{c}%
}{10^{9}\ K}\right)  ^{7}\,C_{A}^{2}F\left(  \tau\right)  \ \frac
{\mathrm{ergs}}{\mathrm{cm}^{3}\mathrm{s}}, \label{E}%
\end{align}
where $M_{p}$ is the bare proton mass, $C_{A}^{2}=g_{A}^{2}\simeq1.6$ (for
neutrons) , and the function $F\left(  \tau\right)  $ is defined as
\begin{align}
F\left(  \tau\right)   &  =\tau^{7}y^{2}\int_{0}^{\infty}dx\frac{\left(
x^{2}+y^{2}\right)  ^{2}}{\left(  e^{\sqrt{x^{2}+y^{2}}}+1\right)  ^{2}%
}\nonumber\\
&  \times\left(  \frac{M^{\ast2}}{M^{2}}\frac{\left(  1+g_{1}/12\right)  ^{2}%
}{\left\vert 1+g_{0}\alpha\left(  x,y\right)  /2\right\vert ^{2}}+\frac
{11}{21}\frac{1}{\left\vert 1+g_{1}\alpha\left(  x,y\right)  /6\right\vert
^{2}}\right)  \label{F}%
\end{align}
The function $\alpha\left(  x,y\right)  $ can be recast as
\begin{align}
\alpha\left(  x,y\right)   &  =-\frac{1}{2}\mathcal{P}\int_{0}^{\infty}%
\frac{d\lambda}{\sqrt{\lambda^{2}+y^{2}}}\frac{y^{2}}{\left(  x^{2}%
-\lambda^{2}\right)  }\tanh\frac{\sqrt{\lambda^{2}+y^{2}}}{2}\nonumber\\
&  +i\frac{\pi}{4}\frac{y^{2}}{x\sqrt{x^{2}+y^{2}}}\tanh\frac{\sqrt
{x^{2}+y^{2}}}{2}. \label{a}%
\end{align}
In numerical estimates, we use the fit expression of the energy gap dependence
on the temperature (see, e.g., Ref. \cite{Kaminker}):
\begin{equation}
y\left(  \tau\right)  =\sqrt{1-\tau}\left(  1.456-\frac{0.157}{\sqrt{\tau}%
}+\frac{1.764}{\tau}\right)  . \label{vfit}%
\end{equation}

\begin{figure}[ptb]
\includegraphics{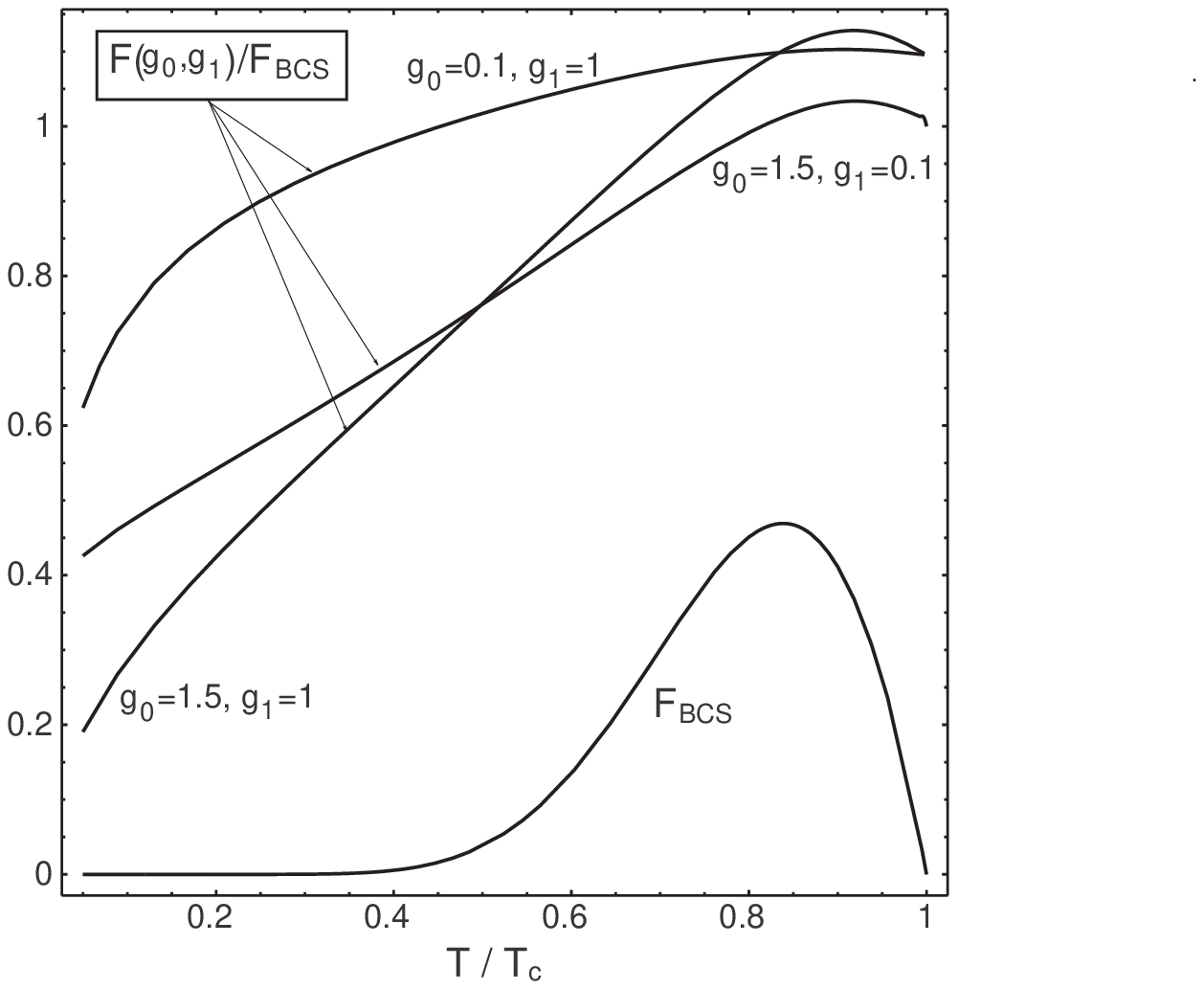}\caption{The temperature dependence of neutrino energy
losses. Lowest curve -- the function $F_{\mathrm{BCS}}$, as given by Eq.
(\ref{FBCS}). Upper curves -- the ratio $F/F_{\mathrm{BCS}}$ for three
different combinations of Landau parameters $g_{0},g_{1}$ shown near the
curves.}%
\label{fig1}%
\end{figure}

Unfortunately, the Landau parameters $g_{0},g_{1}$ are poorly known up to now.
These are known to depend on the baryon density and could be of the order of
unity \cite{Sapershtein}, \cite{Rodin}. Extracted from nuclear data,
$g_{0}=1.5$, while $g_{1}$ is unknown \cite{Migdal}. In our estimate, we use
three different combinations of these parameters. The result of numerical
evaluation is shown in FIG. \ref{fig1}, where we compare the energy losses
according Eq. (\ref{E}) with the BCS expression (\ref{EABCS}), which can be
cast in the same form as Eq. (\ref{E}) but with the function $F\left(
\tau\right)  $ replaced by%
\begin{equation}
F_{\mathrm{BCS}}\left(  \tau\right)  =\tau^{7}y^{2}\int_{0}^{\infty}%
dx\frac{\left(  x^{2}+y^{2}\right)  ^{2}}{\left(  e^{\sqrt{x^{2}+y^{2}}%
}+1\right)  ^{2}}. \label{FBCS}%
\end{equation}
This function is represented by the lowest curve. The upper curves represent
the ratio $F\left(  \tau\right)  /F_{\mathrm{BCS}}\left(  \tau\right)  $ for
three different combinations of the Landau parameters. 

\section{Summary and conclusion}

In this paper, we have investigated the Fermi-liquid effects in the neutrino
emission at the pair recombination of thermal excitations in a superfluid
crust of neutron stars. For this purpose, we have calculated the weak response
functions of superfluid fermion system at finite temperatures while taking
into account the particle-hole interactions near the Fermi surface. For the
calculation, we used Leggett approach to strongly interacting Fermi liquid
with pairing. In the case $\mathrm{q}V_{F}\ll\mathsf{\Delta}$, typical for the
weak processes in the nonrelativistic baryon matter of neutron stars, we have
derived the response functions valid at finite temperature and for arbitrary
transferred energy $\omega\lessgtr\Delta$. Our general expressions, as given
by Eqs. (\ref{K1}), (\ref{T}), (\ref{K00}), and (\ref{Kij}), naturally
reproduce the well-known results \cite{Leggett}, \cite{Larkin} obtained for
the case of small transferred energy, $\omega\ll\Delta,$ as well as the
response functions obtained for arbitrary $\omega$ in the BCS approximation
\cite{L08}.

In the kinematic domain $\omega>2\Delta$ and $\mathrm{q}<\omega$, we have
carefully calculated the imaginary part of the response functions up to the
necessary accuracy, what allows us to evaluate the neutrino energy losses
caused by the pair recombination while taking into account the Fermi-liquid effects.

In the vector channel, we found that the spherical harmonic of the
particle-hole interaction does not affect the imaginary parts of polarization
functions in the time-like domain. The imaginary part of both the longitudinal
and transverse polarization functions are proportional to $V_{F}^{4}$, and
thus the particle-hole interactions are not able to increase substantially the
intensity of neutrino-pair emission through the vector channel.

The imaginary part of the axial polarization is suppressed as $V_{F}^{2}$,
therefore the dominating neutrino emission occurs in the axial channel. We do
not support the statement of Ref. \cite{KV08} that the particle-hole
interactions can be ignored [see the discussion after Eq. (33) of Ref.
\cite{KV08}).\ Our analytic expression (\ref{E}) and numerical estimates
demonstrate the important role of the Fermi liquid effects in the considered process.

\section{HERE}

Discarding the particle-hole interactions means that the result obtained in
Ref. \cite{KV08}, as a matter of fact, corresponds to the BCS approximation.
This approximation has been used before by several authors. Therefore for
comparison, we consider the BCS limit of our Eq. (\ref{EA}) which can be
obtained by putting $g_{0}=g_{1}=0$. The detailed analysis of some
controversial results of different authors can be found at the end of Sec. VI.

For completeness, it is helpful to discuss additionally the case, when the
quasiparticles carry an electric charge. Though the direct neutrino
interaction with recombining protons is screened by the proton background
\cite{LP06}, the proton quantum transitions can excite background electrons,
thus inducing the neutrino-pair emission by the electron plasma. This effect
has been already studied in Refs. \cite{L00}, \cite{L01}; therefore, we only
briefly revisit this problem in the light of modern theory to understand
whether the plasma effects can lead to noticeable neutrino energy losses
through the vector channel. For the sake of simplicity we consider a
degenerate plasma consisting of nonrelativistic superfluid protons and
relativistic electrons. As found in Refs. \cite{L00}, \cite{L01}, the role of
the electron background, in this case, consists of the effective
renormalization of the proton vector weak coupling constant, $c_{V}^{\left(
p\right)  }/2\rightarrow c_{V}^{\left(  e\right)  }$. \ Thus we find that the
electron background strongly increases the effective proton vector weak
coupling with the neutrino field, $\left(  4c_{V}^{\left(  e\right)  }%
/c_{V}^{\left(  p\right)  }\right)  ^{2}\simeq\allowbreak576.$ However, this
huge factor should not mislead the reader, because it arises only as a result
of a very small proton coupling constant, $c_{V}^{\left(  p\right)  }\ll
c_{V}^{\left(  e\right)  }$. Since the degenerate electron plasma can be
considered in the collisionless approximation, the imaginary part of the
medium polarization arises from the proton pair recombination and therefore is
proportional to $V_{F}^{4}$, where $V_{F}\ll1$ is the Fermi velocity of
protons. Thus the neutrino emission through the vector channel is suppressed
by a small factor $V_{F}^{4}$ and may be ignored in comparison with the
dominating neutrino radiation in the axial channel, where the neutrino energy
losses are suppressed as $V_{F}^{2}$.

We now return to the Fermi-liquid effects incorporated in Eq. (\ref{EA}). The
magnitudes of the Landau parameters $g_{0},g_{1}$ are poorly known and depend
on the baryon density. By modern estimates \cite{Sapershtein}, \cite{Rodin},
these could be of the order of unity. Thus the Fermi-liquid effects can
notably modify the emissivity dependence on the temperature and the matter
density as compared to that found in the BCS approximation. This, however,
cannot change the main conclusion that the dominating contribution to the
neutron and proton emissivity comes from the axial channel of weak
interactions \cite{L08}. This means that the neutrino energy losses are to be
suppressed as compared to that of Ref. \cite{FRS76} by a factor of $V_{F}^{2}%
$. This could serve by a natural explanation of the observed superburst
ignition discussed in the Introduction.

\end{document}